\documentclass[conference]{IEEEtran}
\IEEEoverridecommandlockouts

\usepackage{cite}
\usepackage{amsmath,amssymb,amsfonts}
\usepackage[ruled,linesnumbered]{algorithm2e}
\usepackage{graphicx}
\usepackage{makecell}
\usepackage{multicol}
\usepackage{multirow}
\usepackage{textcomp}
\usepackage{xcolor}
\def\BibTeX{{\rm B\kern-.05em{\sc i\kern-.025em b}\kern-.08em
    T\kern-.1667em\lower.7ex\hbox{E}\kern-.125emX}}

\newtheorem{theorem}{Theorem}

\newtheorem{proposition}{Proposition}
\newtheorem{lemma}{Lemma}

\newtheorem{example}{Example}

\begin{document}

\title{Fine-tuning ORBGRAND with Very Few Channel Soft Values\\

\thanks{This work was supported in part by the National Natural Science Foundation of China under Grant 62231022.}
}

\author{%
\IEEEauthorblockN{Li Wan, Huarui Yin, Wenyi Zhang}
\IEEEauthorblockA{Department of Electronic Engineering and Information Science\\
                   University of Science and Technology of China\\
                   Email: wan\_li@mail.ustc.edu.cn, yhr@ustc.edu.cn, wenyizha@ustc.edu.cn}
}

\maketitle

\begin{abstract}
    Guessing random additive noise decoding (GRAND) is a universal decoding paradigm that decodes by repeatedly testing error patterns until identifying a codeword, where the ordering of tests is generated by the received channel values. On one hand, while testing error patterns in a descending order of posterior probabilities leads to maximum likelihood decoding, its implementation complexity is prohibitive. On the other hand, testing error patterns with a prescribed set of error patterns permuted by the ranking among magnitudes of log-likelihood ratios (i.e., ordered reliability bits, ORB) enables efficient implementation, but results in performance loss for finite-length codes. Aiming at harnessing the strengths of these two approaches, this work proposes a fine-tuning method to improve ORBGRAND, adjusting the ordering of tests with the aid of very few exact channel soft values. This method is based on a metric for assessing the ``well-orderedness'' of error patterns. The metric is studied via the lens of the asymptotic theory of integer partitioning, which provides highly accurate estimation in numerical experiments. The metric then leads to an effective identification of fine-tuning to conduct, at the cost of a negligible increment of complexity. Numerical experiments demonstrate that the proposed fine-tuning method achieves a substantial performance enhancement compared with ORBGRAND.
\end{abstract}


\section{Introduction}
Guessing random additive noise decoding (GRAND) \cite{duffy2019capacity} is a recently proposed universal decoding paradigm, suitable for codes with short codeword length and high code rate. These attributes make GRAND a promising candidate for integration into emerging communication scenarios that demand exceptional reliability and low latency, such as ultra-reliable low-latency communication (URLLC) \cite{yue2023efficient} \cite{rowshan2024channel}.

For channels with soft output values, GRAND leverages the reliability of channel outputs to repeatedly test error patterns until identifying a codeword when flipping the hard-decision channel output vector using an error pattern. The key aspect of a GRAND algorithm, therefore, is how the error patterns are generated. The soft GRAND (SGRAND) algorithm \cite{solomon2020soft} can achieve maximum likelihood (ML) decoding by sorting error patterns in a descending order of posterior probabilities \cite{liu2022orbgrand} \cite[Ch. 10]{Lin2004ErrorCC}. Its implementation, however, is prohibitively complex and there is currently no feasible parallel algorithm. 
A key simplification involves using only the ranking of log-likelihood ratio (LLR) magnitudes, leading to the ordered reliability bits GRAND (ORBGRAND) algorithm \cite{duffy2022ordered}, whose error patterns can then be obtained by a basis list of error patterns generated offline in advance, permuted by the ORB. This feature leads to low-complexity and parallelizable hardware implementation \cite{abbas2022high} \cite{ji2024efficient}.

To further reduce complexity or adapt to specific channels, subsequent studies improved ORBGRAND by designing error pattern lists based on channel characteristics \cite{condo2021high} \cite{liu2022orbgrand} \cite{li2023pre} \cite{wan2024approaching} \cite{ji2024efficient}. These algorithms retain the idea of permute error patterns using the order of LLRs and the feature of low complexity implementation, so they are classified as ORB-type GRAND algorithms. By reasonably arranging error patterns, such ORB-type GRAND algorithms can achieve block error rate (BLER) performance that is fairly close to the ML bound \cite{wan2024approaching}.

It has been observed that since ORB-type GRAND algorithms utilize the ordering information of LLRs instead of their exact values for decoding, there is an inevitable loss of information that leads to reduced performance. To address this issue, this paper proposes a fine-tuning ORBGRAND algorithm that incorporates received soft values. This method maintains the efficiency of ORB-type GRAND while approaching the performance of ML decoding using only 1 or 2 carefully selected soft values.

Our method utilizes the number of inversion pairs as a metric to assess the ``well-orderedness'' of error patterns, aiding in the determination of optimal soft values for decoding. We formulate an optimization problem to base on this metric. The integer partition theory is applied to efficiently solve the optimization problem, enabling the effective selection of channel soft values during the decoding process.

Based on the analysis of error pattern structure under partition theory, this paper proposes an algorithm to fine-tune the original error pattern of ORB-type GRAND, and obtain a new sequence using essentially linear complexity, avoiding real-time generation of new error patterns. Simulation results show that generating the fine-tuned error patterns incurs low computational cost. Compared to the original ORB-type GRAND, the resulting decoding algorithm requires fewer guesses and achieves lower BLER, with a gap to within 0.05dB of SGRAND.

By using a portion of exact channel soft values, the proposed fine-tuning ORBGRAND acts as an intermediate form between SGRAND and ORBGRAND, and using only a few exact values achieves both ORBGRAND's efficient implementation and SGRAND's superior performance.


\section{Preliminaries}

\subsection{Transmission Model}\label{2.1}

Consider a block coding system where information bit $\underline{U} \in \mathbb{F}_{2}^{K}$ are encoded into codewords $\underline{W} \in \mathcal{C} \subseteq \mathbb{F}_{2}^{N}$ with code rate $R = K/N$. For simplicity, we consider antipodal signaling over an additive white Gaussian noise (AWGN) channel, leading to transmitted vector $\underline{X}$ satisfying $X_i = 1$ if $W_i = 0$ and $X_i = -1$ if $W_i = 1$, $i = 1, \ldots, N$. The resulting AWGN channel output vector satisfies $\underline{Y}|\underline{X} \sim \mathcal{N}(\underline{X}, \sigma^2 \mathbf{I}_{N \times N})$. Given $\underline{Y}$, the LLRs are derived as:
\begin{equation*}\label{eq:LLR}
    \text{LLR}_{i} = \log\frac{P(Y_i \mid W_{i} = 0)}{P(Y_i \mid W_{i} = 1)} = \frac{2}{\sigma^2} Y_i, i = 1, \ldots, N,
\end{equation*}
we use $L_i = |\text{LLR}_i|$ to represent the amplitude of LLR, that is, the reliability of this symbol, and use $\ell_i$ to represent the realization of $L_i$. The reliability measure $L$ follows a folded normal distribution with probability density function (PDF):
\begin{equation}\label{eq:pdf_ell}
\frac{\sigma}{2\sqrt{2\pi}} \left[ \exp\left(-\frac{\sigma^2 (\ell - \frac{2}{\sigma^2})^2}{8}\right) + \exp\left(-\frac{\sigma^2 (\ell + \frac{2}{\sigma^2})^2}{8}\right) \right]
\end{equation}
for $\ell \geq 0$ and $f_{L}(\ell) = 0$ for $\ell < 0$.

We introduce the hard decision function $\theta(\cdot)$ as $\theta(y) = 0$ if $y \geq 0$ and $1$ otherwise. For $\underline{Y}$, denote the vector of $[\theta(Y_1), \ldots, \theta(Y_n)] \in \mathbb{F}_2^N$ by $\theta(\underline{Y})$ for simplicity.

\subsection{GRAND and Metric of Well-orderedness}\label{2.2}
The GRAND decoding paradigm operates through the following procedure:
\begin{itemize}
    \item Generate ordered error patterns $\{\underline{e}(t)\}_{t=1}^{2^N}$, each $\underline{e}\in \mathbb{F}_2^N$.
    \item For each pattern $\underline{e}(t)$, compute $\theta(\underline{y}) \oplus \underline{e}(t)$.
    \item Terminate when $\theta(\underline{y}) \oplus \underline{e} \in \mathcal{C}$ or maximum trials $T$ reached.
\end{itemize}


For clarity, denote the ordered list of error patterns as $\mathcal{E} = \{\underline{e}(1), \ldots, \underline{e}(2^N)\}$, and the algorithm tests only the first $T$ elements of $\mathcal{E}$. The way of generating the error pattern set $\mathcal{E}$ determines the performance of a GRAND algorithm. The SGRAND algorithm \cite{solomon2020soft}
implements ML decoding by ordering error patterns according to:
\begin{equation}
    \sum_{i=1}^{N}\ell_i e_i(t) \leq \sum_{i=1}^{N}\ell_i e_i(t'), \  \forall \ 1\leq t \leq t' \leq 2^N. \label{eq:ML}
\end{equation}
That is, we can define $\zeta(\underline{e}(t))= \sum_{i=1}^{N}\ell_i e_i(t)$, $t = 1, \ldots, 2^N$, then $\{\zeta(\underline{e}(t))\}_{t=1}^{2^N}$ are arranged in an ascending order.

Instead, the ORBGRAND algorithm uses the ranking of $\{\ell_1,...,\ell_N\}$ to replace their actual values when generating $\mathcal{E}$ \cite{duffy2022ordered}, and this idea can be further extended to the so-called ORB-type GRAND algorithms \cite{wan2024approaching}. An ORB-type GRAND algorithm has a basis error pattern set $\Tilde{\mathcal{E}}=\{\Tilde{\underline{e}}(1),...,\Tilde{\underline{e}}(2^N)\}$, which, in most cases, is arranged in a way that satisfies:\footnote{There are some error pattern sets $\mathcal{E}$ that are not generated using this criterion, and this paper will not fine-tune such structures.}
\begin{equation}\label{eqn:ORB-type-criterion}
    \sum_{i = 1}^N \gamma_i \Tilde{e}_i(t) \leq \sum_{i = 1}^N \gamma_i \Tilde{e}_i(t'), \ \forall \ 1 \leq t \leq t' \leq 2^N,
\end{equation}
where $\underline{\gamma}$ does not depend upon $\underline{y}$ and hence $\Tilde{\mathcal{E}}$ can be prepared offline in advance. For ORBGRAND, we simply let $\underline{\gamma} = [1,2,\ldots,N]$, and there also exist other choices of $\underline{\gamma}$  leading to a variety of ORB-type GRAND algorithms \cite{duffy2022ordered} \cite{liu2022orbgrand} \cite{wan2024approaching}. Upon receiving $\underline{y}$, calculate the ranking vector $\underline{r}$ in which $r_i$ is the rank of $\ell_i$ among $\{\ell_1,...,\ell_N\}$, in an ascending order. With $\underline{r}$, the ordered list of error patterns $\mathcal{E}$ is obtained as $\underline{e}(t) = [\Tilde{e}_{r_1}(t),...,\Tilde{e}_{r_N}(t)]$, for $t = 1, \ldots, 2^N$. From (\ref{eqn:ORB-type-criterion}) it then immediately follows that
\begin{equation}\label{eq:orb}
    \sum_{i=1}^{N}\gamma_{r_i} e_i(t) \leq \sum_{i=1}^{N}\gamma_{r_i} e_i(t'), \  \forall \ 1\leq t \leq t' \leq 2^N. 
\end{equation}
\noindent Let us define $\Gamma(\underline{e}(t))= \sum_{i=1}^{N}\gamma_{r_i} e_i(t)$, which will be used subsequently.

Comparing the SGRAND (i.e., ML) criterion (\ref{eq:ML}) and the ORB-type GRAND criterion (\ref{eq:orb}), we see that the suboptimality of ORB-type GRAND algorithms stems from the fact that, with $\mathcal{E}$ obtained according to (\ref{eq:orb}), $\zeta(\underline{e}(t))$, $t = 1, \ldots, T$, are not necessarily arranged in an ascending order. We are therefore motivated to propose the following metric for assessing the ``well-orderedness'' of an ORB-type GRAND algorithm:
\begin{align}
    & \eta_{\text{GRAND}}(\underline{y}) = \frac{\sum_{1\leq t<t'\leq T}\mathbf{1}(\zeta(\underline{e}(t))>\zeta(\underline{e}(t')))}{T(T-1)/2}, \label{eq:eta}\\
    & \eta_{\text{GRAND}} = \mathbf{E}_{\underline{Y}}(\eta_{\text{GRAND}}(\underline{y})), \label{eq:e_eta}
\end{align}
in which $\sum_{1\leq t<t'\leq T}\mathbf{1}(\zeta(\underline{e}(t))>\zeta(\underline{e}(t')))$ is the number of ``reverse pairs'' in all pairs of $\mathcal{E}$. The subscripts are the different algorithms used. In particular, for the SGRAND algorithm, we have $\eta_{\text{SGRAND}} = 0$.

\section{Main Results}

The subsection \ref{subsec:3.1} presents an optimization problem for selecting soft value locations, in order to select the positions that produce the most inversion pairs. Subsection \ref{subsec:3.2} and \ref{subsec:3.3} analyze this counting problem. Subsection \ref{subsec:3.4} discusses the fine-tuning algorithm, and \ref{subsec:3.5} evaluates the complexity of the algorithm.

\subsection{Problem Statement}\label{subsec:3.1}

Given a received vector $\underline{y}$, using its ranking vector $\underline{r}$ only leads to ORB-type GRAND algorithms. Now we want to replace a few positions in $\underline{r}$ by their exact magnitudes of LLRs. Intuitively, these positions should be ``critical'' to have a large impact on the metric of well-orderedness.

%


Let $\mathcal{D}=\{d_1,..,d_{D}\} \subseteq \{1,2,...,N\}$ indicate $D$ positions in $\underline{y}$. After calculating the ranking vector $\underline{r}$ from $\underline{y}$, define $\Tilde{\underline{\ell}}$ as: $\Tilde{\ell}_i = \ell_i$ if $i\in \mathcal{D}$; otherwise, $\Tilde{\ell}_i = \gamma_{r_i}$. Our first task is to identify the set $\mathcal{D}$ that maximizes the number of reverse pairs under $\Tilde{\zeta}(\underline{e}(t)) = \sum_{i=1}^{N}\Tilde{\ell}_ie_i$, i.e.,
\begin{equation}\label{eq:problem}
    \mathcal{D}^{*} = \arg\max_{d_1,d_2,...,d_D}\sum_{1\leq t < t' \leq T}\mathbf{1}\left(\Tilde{\zeta}(\underline{e}(t)) > \Tilde{\zeta}(\underline{e}(t')\right).
\end{equation}
The heuristic behind (\ref{eq:problem}) is that, the channel soft values at the positions of $\mathcal{D}^{*}$ in $\underline{y}$ lead to the largest number of reverse pairs, and hence these positions need be fine-tuned.

\subsection{Decomposition of Reverse Pairs}\label{subsec:3.2}

Given any $\mathcal{D}$, according to the error pattern on its corresponding $D$ positions in $\underline{y}$, $\mathcal{E}$ can be divided into $2^{D}$ subsets as $\mathcal{E} =  \bigcup_{u=0}^{2^D-1}\mathcal{E}_u$ with
\begin{equation}\label{eq:def_of_subset}
    \mathcal{E}_u = \{\underline{e}\in \mathbb{F}_2^N \mid \forall\ j \in \{1,2,...,D\},e_{d_j} = \text{dec2bin}^{(D)}_{j}(u)\},
\end{equation}
where $\text{dec2bin}^{(D)}_{j}(u)$ denotes the $j$-th digit obtained by converting $u$ from decimal representation to $D$-bit binary representation. Each error pattern $\underline{e}$ belongs to precisely one $\mathcal{E}_u$, and possesses the following characteristics:
\begin{proposition}\label{prop:1}
For any $\underline{e}\in \mathcal{E}_u$, its $\Tilde{\zeta}(\underline{e})$ is given by:
\begin{equation*}
    \Tilde{\zeta}(\underline{e}) = \Gamma(\underline{e}) + \delta(u),
\end{equation*}
in which we define $\delta(u) = \sum_{i\in \mathcal{D}}(\ell_i-\gamma_{r_i}) (\text{dec2bin}^{(|D|)}_{j}(u))$.
\end{proposition}
\begin{IEEEproof}
From the definition of $\Tilde{\zeta}(\underline{e})$, we can directly get
\begin{align}
    \Tilde{\zeta}(\underline{e}) & = \sum_{i=1}^{N}(\Tilde{\ell}_i - \gamma_{r_i} + \gamma_{r_i})e_i \notag  \\
    & = \sum_{i=1}^{N}\gamma_{r_i}e_i + \sum_{i\in \mathcal{D}}(\Tilde{\ell}_i - \gamma_{r_i})e_i + \sum_{i\notin \mathcal{D}}(\Tilde{\ell}_i - \gamma_{r_i})e_i, \label{eq:prop_1_1}
\end{align}
where the first term in \eqref{eq:prop_1_1} is $\Gamma(\underline{e})$, the third term is zero because for $i\notin\mathcal{D}$ we have $\Tilde{\ell}_i = \gamma_{r_i}$, and the second term can be rewritten as $\delta(u)$ according to \eqref{eq:def_of_subset}.
\end{IEEEproof}

The following theorem provides an equivalent form of the number of reverse pairs in \eqref{eq:problem}:
\begin{theorem}\label{th:1}
    Denote by $\mathcal{I}$ the objective function in \eqref{eq:problem}, i.e., the number of reverse pairs under $\Tilde{\zeta}$. Then $\mathcal{I}$ can be expressed as $\mathcal{I} = \sum_{0\leq u ,v \leq 2^D-1} \mathcal{I}_{u,v}$ with $\mathcal{I}_{u,v}$ equal to:
\begin{align}
    \sum_{\substack{\underline{e}(t)\in \mathcal{E}_u,\underline{e}(t')\in \mathcal{E}_v\\1 \leq t < t' \leq T}} \mathbf{1}(\Gamma(\underline{e}(t)) + \delta(u) > \Gamma(\underline{e}(t')) + \delta(v)). \label{eq:th_1}
\end{align}
    
\end{theorem}

\begin{IEEEproof}
Starting from the expression of the objective function in \eqref{eq:problem}, we get:
\begin{align}
    \mathcal{I} & = \sum_{1\leq t < t'\leq T}\mathbf{1}(t < t')\cdot \mathbf{1}(\Tilde{\zeta}(\underline{e}(t)>\Tilde{\zeta}(\underline{e}(t'))) \notag  \\
    & = \sum_{1\leq t < t'\leq T}\left(\sum_{0\leq u,v\leq 2^D-1}\mathbf{1}(\underline{e}(t)\in \mathcal{E}_u \text{ and }\underline{e}(t')\in \mathcal{E}_v)\right) \notag \\
    & \quad \quad \quad \cdot \mathbf{1}(\Tilde{\zeta}(\underline{e}(t)>\Tilde{\zeta}(\underline{e}(t'))) \label{eq:inverse_1}\\
    & = \sum_{0\leq u,v\leq 2^D-1}\sum_{\underline{e}(t)\in \mathcal{E}_u,\underline{e}(t')\in \mathcal{E}_v}\mathbf{1}(1 \leq t < t'\leq T ) \notag \\
    & \quad \quad \quad \cdot\mathbf{1}(\Gamma(\underline{e}(t)) + \delta(u) >\Gamma(\underline{e}(t')) + \delta(v)), \label{eq:inverse_2}
\end{align}

\noindent where \eqref{eq:inverse_1} is simply $\sum_{0\leq u,v\leq 2^D-1}\mathbf{1}(\underline{e}(t)\in \mathcal{E}_u \text{ and }\underline{e}(t')\in \mathcal{E}_v) = 1$ because $\{\mathcal{E}_{u}\}_{u=0}^{2^{D}-1}$ constitute a partition of $\mathcal{E}$. Substituting the result of Proposition \ref{prop:1} into \eqref{eq:inverse_2} thus completes the proof.
\end{IEEEproof}

Theorem \ref{th:1} suggests that, since $\Gamma(\cdot)$ depends on $\mathcal{E}$ of the ORB-type GRAND algorithm, the choice of $\mathcal{D}$ only influences $\delta(\cdot)$. 
Next we exploit an asymptotic property of $\Gamma(\cdot)$ for the estimation of $\mathcal{I}_{u,v}$ and $\mathcal{I}$.

\subsection{Estimation of Number of Reverse Pairs}\label{subsec:3.3}
According to the ORB-type GRAND criterion \eqref{eq:orb}, for a given error pattern $\underline{e}$, its position in $\mathcal{E}$ can be expressed as follows:
\begin{equation}\label{eq:o_function}
    \mathcal{O}(\Gamma(\underline{e})) = \sum_{t = 1}^{2^N}\mathbf{1}\left( \sum_{i = 1}^{N}\gamma_{r_i}  e_i(t) \leq  \Gamma(\underline{e})\right).
\end{equation}
We then propose to approximate the value of $\mathcal{O}(\Gamma(\underline{e}))$ using a result in integer partition:
\begin{lemma}[Szekeres 1987 \cite{SzekeresAsymptotic}]
    For an integer $n$, the number of its non-repeating partitions, $q(n)$, asymptotically behaves like 
    \begin{equation}
        Q(n) = \frac{e^{\pi \sqrt{n / 3}}}{4 \cdot 3^{1 / 4} n^{3 / 4}}, \label{eq:lemma}
    \end{equation}
    in the sense that the the ratio between $q(n)$ and $Q(n)$ tends to one as $n \rightarrow \infty$.
\end{lemma}

For ORBGRAND which have $\underline{\gamma} = [1, 2, ..., N]$, and whenever $m \leq N$, $\mathcal{O}(m)$ is exactly the sum of $q(n)$ for $n = 1, \ldots, m$. In practical implementation of ORBGRAND, for any reasonable codeword length $N$, the value of $\mathcal{O}(N)$ is already exceedingly large and there is no need to consider $m > N$. So the following estimate of $\mathcal{O}(m)$ will be of interest for our purposes:
\begin{align}
    \mathcal{O}(m) & \approx \sum_{n=1}^{m}\frac{e^{\pi \sqrt{n / 3}}}{4 \cdot 3^{1 / 4} n^{3 / 4}} \approx
    \int_{0}^{m} \frac{e^{\pi \sqrt{x / 3}}}{4 \cdot 3^{1 / 4} x^{3 / 4}}\text{d}x \notag \\
    & =\frac{1}{2}\mathrm{erfi}\left(\frac{\sqrt{\pi}m^{1/4}}{3^{1/4}} \right) := \Tilde{\mathcal{O}}(m), \label{eq:fit2}
\end{align}
in which $\mathrm{erfi}(x) = \frac{2}{\sqrt{\pi}}\int_{0}^{x}e^{z^2}\text{d}z$ is the imaginary error function. For general ORB-type GRAND algorithms, $\underline{\gamma}$ are not necessarily distinctly integer-valued, but we can still follow the integral form in \eqref{eq:lemma} and use $a\cdot e^{(b\cdot m^c)}/m^{d}$ to fit $\Tilde{\mathcal{O}}'(m)$, where the $[a,b,c,d]$ are parameters to fit the $\mathcal{O}(m)$. In Section \ref{sec:performance} it will be shown via numerical experiments that $\Tilde{\mathcal{O}}(m)$ provides a highly accurate estimate of $\mathcal{O}(m)$.

Based upon $\Tilde{\mathcal{O}}(m)$, we propose the following estimate of $\mathcal{I}$. It is obtained by manipulating \eqref{eq:inverse_2} and replacing discrete sums by integrals. The detailed derivation steps are omitted. We emphasize that the estimate is not meant to be rigorous and will be only used for developing the fine-tuning method in the next subsection. Its effectiveness will be validated via numerical experiments in Section \ref{sec:performance}.
\begin{align}
    &\mathcal{I} \approx \sum_{0\leq u<v\leq 2^D-1} p_u(T) p_v(T) |\delta(u,v)|\cdot \notag \\
    &\quad\int_{0}^{\Tilde{\mathcal{O}}^{-1}(T) - \left|\frac{\delta(u,v)}{2}\right|}\Tilde{\mathcal{O}}'(m) \cdot \Tilde{\mathcal{O}}'\left(m + \left|\frac{\delta(u,v)}{2}\right|\right)\text{d}m, \label{eq:esti_I}
\end{align}

\noindent in which $p_u(T)$ represents the probability that the error patterns in $\{\underline{e}(1),...,\underline{e}(T)\}$ belong to $\mathcal{E}_u$, can be defined as 
\begin{equation}\label{eq:pT}
    p_u(T) = \frac{1}{T}\sum_{t=1}^{T} \mathbf{1}(\underline{e}(t) \in \mathcal{E}_u),
\end{equation}
and $\delta(u,v)$ represents $\delta(u)- \delta(v)$ in short.

\subsection{Fine-tuning Method}\label{subsec:3.4}
Based on \eqref{eq:esti_I}, we can solve problem \eqref{eq:problem}, obtain the $\mathcal{D}^*$ by maximizing $\mathcal{I}$ over $\mathcal{D}$. Next, we estimate where an error pattern should be moved under $\Tilde{\zeta}$, as given by the following theorem:
\begin{theorem}\label{thm:adjustment}
    For an error pattern $\underline{e}\in \mathcal{E}_u$, when we rearrange error patterns with $\Tilde{\zeta}$, its position should be:
    \begin{equation}\label{eq:new_posi}
        \sum_{0\leq v\leq 2^D-1}p_v(\mathcal{O}(\Gamma(\underline{e})+ \delta(u,v)))\mathcal{O}(\Gamma(\underline{e}) + \delta(u,v)).
    \end{equation}
\end{theorem}
\begin{IEEEproof}
When rearranging error patterns with $\Tilde{\zeta}$, the position of $\underline{e}$ is $\sum_{t = 1}^{2^N}\mathbf{1}\left( \Tilde{\zeta}(\underline{e}(t))\leq  \Tilde{\zeta}(\underline{e})\right)$. Applying Proposition \ref{prop:1}, the position can be rewritten as:
\begin{equation*}
    \sum_{0\leq v\leq 2^D-1}\sum_{\underline{e}(t)\in \mathcal{E}_v}\mathbf{1}\left( \Gamma(\underline{e}(t)) \leq  \Gamma(\underline{e}) + \delta(u,v)\right).
\end{equation*}
Plugging into $p_v(\cdot)$ in \eqref{eq:pT} hence completes the proof.
\end{IEEEproof}

Based on the above analysis, the Algorithm \ref{algo:1} provides an outline of an ORB-type GRAND algorithm with the proposed fine-tuning method. In the line 2 of the algorithm, we further simplify the solution to \eqref{eq:problem}, since the integral in \eqref{eq:esti_I} is still of high complexity, although it is accurate enough:
\begin{equation}\label{eq:obtain_D}
    (d_1^{*},...,d_D^{*}) = \arg\max_{{d_1,...,d_D}}\sum_{u<v} p_u(T) p_v(T) |\delta(u,v)|.
\end{equation}
This simplification is based on our empirical observation that the integral in \eqref{eq:esti_I} is insensitive to the value of $\delta(u,v)$, as shown in the simulation results in \ref{subsec:4.2}. We give two examples to show that the calculation is simple enough when $D$ is small:
\begin{example}\label{ex:1}
    When $D=1$, \eqref{eq:obtain_D} becomes
    \begin{equation}\label{eq:d_eq_1}
        d_1^{*} = \arg\max_{d_1} p_0(T) (1-p_0(T)) \left|\ell_{d_1} - \gamma_{r_{d_1}}\right|.
    \end{equation}
    Where $p_0(T)$ is the probability of $e_{d_1}(t) = 0$ in the first $T$ error patterns, then $p_1(T) = 1 - p_0(T)$. After knowing $r_{d_1}$, $p_0(T)$ is a value that can be calculated in advance and depends on the error pattern list $\Tilde{\mathcal{E}}$. And $|\delta(0,1)| = \left|\ell_{d_1} - \gamma_{r_{d_1}}\right|$.
\end{example}

\begin{example}\label{ex:2}
    When $D=2$, \eqref{eq:obtain_D} becomes
        \begin{equation*}
        (d_1^{*},d_2^{*}) = \arg\max_{d_1,d_2}\sum_{0\leq u < v \leq 3} p_u(T) p_v(T) |\delta(u,v)|, \ \text{with}
    \end{equation*}
    \begin{equation*}
    \delta(u,v) = \left\{\begin{array}{cc}
          \gamma_{r_{d_2}}-\ell_{d_2} & (u,v) = (0,1) \\
          \gamma_{r_{d_1}}-\ell_{d_1} & (u,v) = (0,2) \\
          \gamma_{r_{d_2}}-\ell_{d_2} + (\gamma_{r_{d_1}}-\ell_{d_1})  & (u,v) = (0,3) \\
          \gamma_{r_{d_2}}-\ell_{d_2} - (\gamma_{r_{d_1}}-\ell_{d_1}) & (u,v) = (1,2) \\
          \gamma_{r_{d_1}}-\ell_{d_1} & (u,v) = (1,3) \\
          \gamma_{r_{d_2}}-\ell_{d_2} & (u,v) = (2,3) \\
    \end{array}\right..
    \end{equation*}
\end{example}

%
%

\begin{algorithm}[htbp]
\caption{Fine-tuning ORB-type GRAND}\label{algo:1}
\KwData{LLR $\underline{\ell}$, $\underline{\gamma}$, Basis error patterns $\Tilde{\mathcal{E}}$}
\KwResult{$\hat{\underline{w}}$}
$\underline{\ell} \rightarrow (\theta(\underline{y}), \underline{r} ) $, $(\Tilde{\mathcal{E}}, \ \underline{r})\rightarrow \mathcal{E} $  \\
$(d_1^{*},...,d_D^{*}) = \arg\max_{{d_1,...,d_D}}\sum_{u<v} p_u p_v |\delta(u,v)|$\\
$A \leftarrow $ zero array, $idx = 1$ \\
\For{$t\leftarrow 1$ \KwTo $T$}{
    \While{$A(t) = 0$}{
        $u \leftarrow \text{bin2dec}([e_{d_1^{*}}(idx),...,e_{d_D^{*}}(idx)])$\\
        $t_{FT} \leftarrow$ \eqref{eq:new_posi} \\
        $A( t_{FT}  ) = idx$, $idx = idx + 1$ \\
    }
    \If{$\theta(\underline{y}) \oplus \underline{e}(A(t))\in \mathcal{C}$}{
        $\hat{\underline{w}} \leftarrow  \theta(\underline{y}) \oplus \underline{e}(A(t))$\\
        break
    }
}
\end{algorithm}

In Algorithm \ref{algo:1}, we conduct fine-tuning during the execution of the ORB-type GRAND algorithm. To do this, we use \eqref{eq:new_posi} to calculate the fine-tuned position $t_{FT}$ of the error pattern $\underline{e}(idx)$ in $\mathcal{E}$, for $idx$ from $1$ to $T$, and then let $A(t_{FT})=idx$, which means that at the $t_{FT}$-th test, the error pattern is $\underline{e}(idx)$. If $A(t) = 0$, this means that the $\underline{e}$ that should be adjusted to this position has not been found, and we then continue to increase $idx$ until this $A(t)$ is filled. Otherwise, according to the value of $A(t)$, the error pattern to test then is $\underline{e}(A(t))$. 

\begin{example}\label{ex:3}
    Figure \ref{fig:toy_model} illustrates a toy model under $D=1$. A grey box represents an error pattern belonging to $\mathcal{E}_0$, which is adjusted to move backward, while the white boxes belong to $\mathcal{E}_1$. The figure depicts the calculation process at $t=4$: since $A(4)=0$ at this point, we compute the new positions of $\underline{e}(4)$, $\underline{e}(5)$, and $\underline{e}(6)$ within the original error patterns until we find $A(4)=6$, at which point decoding continues.
\begin{figure}[htbp]
    \centering
    \includegraphics[width=0.42\textwidth]{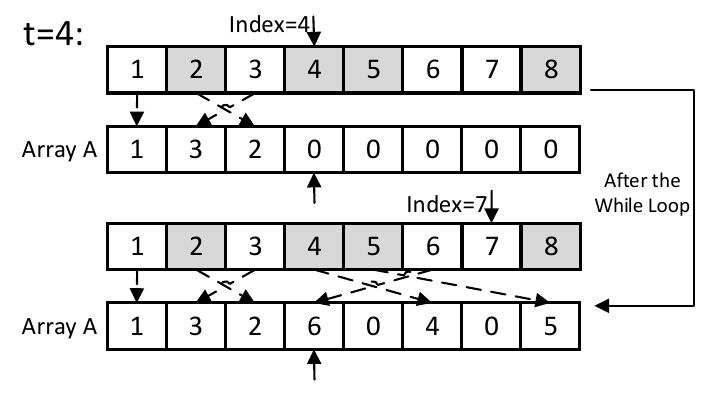}
    \caption{At $t=4$, the loop is repeated three times, and the adjusted positions of $\underline{e}(4)$, $\underline{e}(5)$, and $\underline{e}(6)$ are calculated.}
    \label{fig:toy_model}
\end{figure}
\end{example}
\subsection{Complexity Analysis}\label{subsec:3.5}
In Table \ref{tab:1}, we compare the computational complexity of ORB-type GRAND, SGRAND and fine-tuning ORB-type GRAND in the decoding stage. Before decoding, all algorithms have a ranking complexity of $O(N\log N)$. And with fine-tuning, \eqref{eq:obtain_D} needs to be solved for obtaining $(d_1^{*},...,d_D^{*})$, which incurs a small amount of overhead since $D$ is typically as small as $1$ or $2$ in practice. But for SGRAND, the cost of generating error patterns typically incurs a large number of floating-number operations and is dominating.

Compared with ORB-type GRAND, the additional complexity in the fine-tuning ORB-type GRAND is the while loop (Line 5-9) in Algorithm \ref{algo:1}, which requires determining the set $\mathcal{E}_u$ of the $\underline{e}$ and calculating the fine-tuned position. On average, the number of loops required at time $t$ is about $1$ time: multiple calculations may be required, or there may be no loops (in Example \ref{ex:3}, when $t=6$, $A(6)=4$ has already been calculated). 

\begin{table}[htbp]
\renewcommand{\arraystretch}{1.0}
\centering
\caption{Computational complexity of decoding algorithms}
\label{tab:1}
\small
\begin{tabular}{|c|c|c|c|c|}
\hline
                           &            & At Time $t$      &    Worst Case  \\ \hline
\multirow{2}{*}{\makecell{ORB-type\\GRAND}}  & Generate & /            & /        \\ \cline{2-4}
                           & Test     & $N(N-K)$     & $N(N-K)T$      \\ \hline
\multirow{3}{*}{SGRAND}    & Generate & $2N$ (floating) & $2NT$ (floating)      \\ \cline{2-4} 
                           & Find  & $2\log t$    & $2\log (T!)$ \\ \cline{2-4} 
                           & Test     & $N(N-K)$     & $N(N-K)T$      \\ \hline
\multirow{3}{*}{\makecell{Fine-tuning\\ORB-type \\GRAND}}& Generate & /            & /     \\ \cline{2-4} 
                           & Adjust   & $O(1)$& $O(T)$ \\ \cline{2-4} 
                           & Test     & $N(N-K)$     & $N(N-K)T$      \\ \hline                          
\end{tabular}
\end{table}

\section{Performance Evaluation}\label{sec:performance}

We select ORBGRAND and CDF-ORBGRAND \cite{liu2022orbgrand} \cite{duffy2022ordered} as the baseline ORB-type GRAND algorithms. These two algorithms respectively have $\underline{\gamma} = [1,2,...,N]$ and $\underline{\gamma} =\left[ \text{CDF}^{-1}_{\ell}(\frac{1}{N+1}),...,\text{CDF}^{-1}_{\ell}(\frac{N}{N+1})\right ]$, where $\text{CDF}^{-1}_{\ell}(\cdot)$ is the inverse function of the cumulative distribution function (CDF) of the $L$. We consider the BCH(127,113) code on the AWGN channel, and set $T=10^4$ as the largest number of tests.

\subsection{Estimation of Partition Number}\label{subsec:4.1}

\begin{figure}[htbp]
    \centering
    \includegraphics[width=0.45\textwidth]{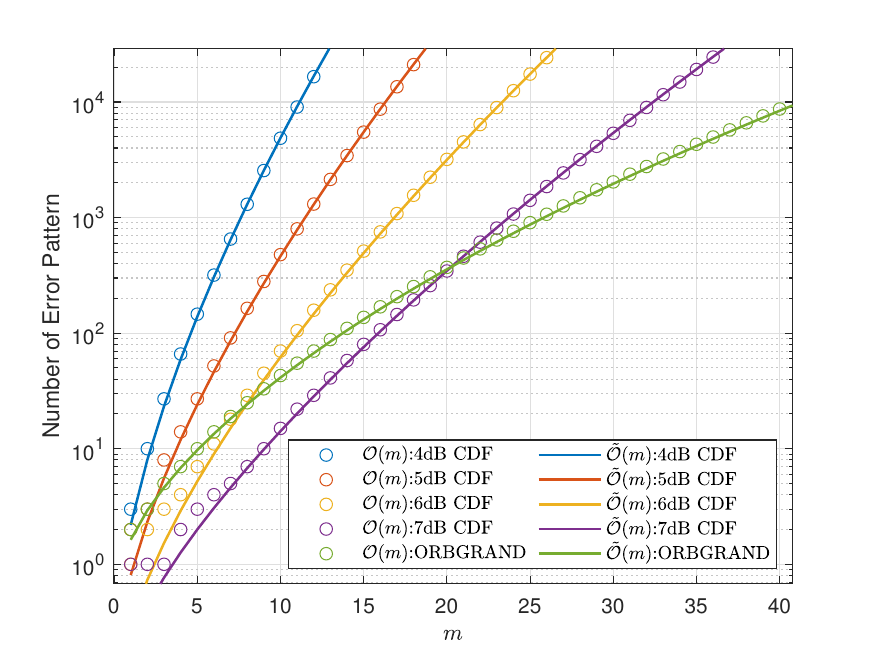}
    \caption{Estimation of $\mathcal{O}(m)$, the fitted curve is very close to the exact value.}
    \label{fig:esti_of_O}
\end{figure}

Figure \ref{fig:esti_of_O} shows the numerical accuracy of $\Tilde{\mathcal{O}}(m)$. From $\underline{\gamma}$, $\mathcal{O}(m)$ can be obtained from $\eqref{eq:o_function}$, shown in discrete circles. Since the CDF of $L$ depends on the channel signal-to-noise ratio, can be obtained from \eqref{eq:pdf_ell}, we plot the $\mathcal{O}(m)$ of CDF-ORBGRAND at Eb/N0 $\in \{4,5,6,7\}$ dB. For ORBGRAND, we approximate $\Tilde{\mathcal{O}}(m)$ via \eqref{eq:fit2}; for CDF-ORBGRAND, we use $a\cdot e^{(b\cdot m^c)}/m^{d}$ to fit $\Tilde{\mathcal{O}}'(m)$. We observe from Figure \ref{fig:esti_of_O} that except for small $m$ at high Eb/N0, $\Tilde{\mathcal{O}}(m)$ is a very accurate estimate of $\mathcal{O}(m)$.

\begin{figure}[htbp]
    \centering
    \includegraphics[width=0.45\textwidth]{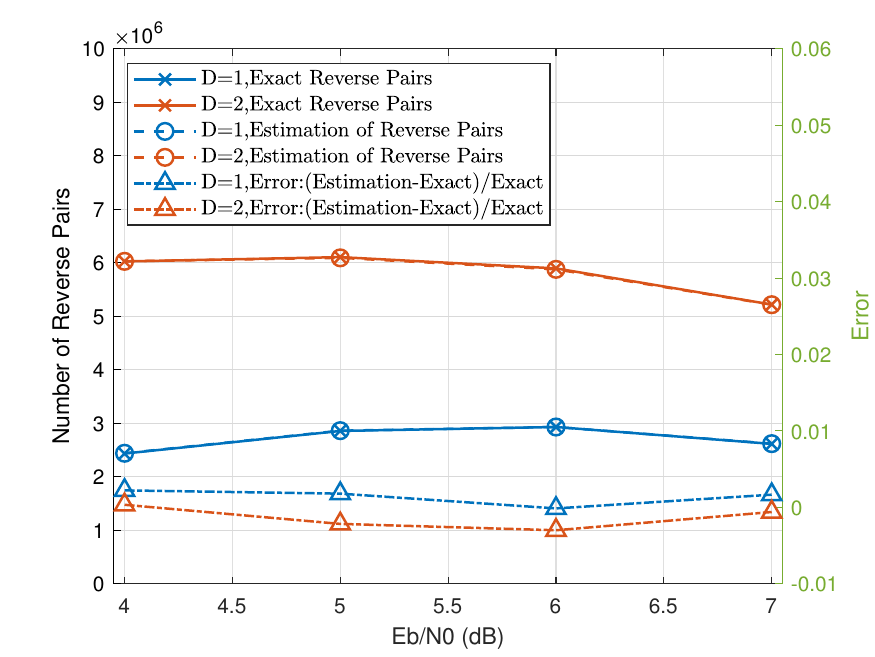}
    \caption{The exact and estimated values of $\mathcal{I}$ are plotted as crosses (``x'') and circles (``o''), respectively. The triangle represents the relative error, and the right axis shows that the relative error is much less than 1\%.}
    \label{fig:reverse_number}
\end{figure}

In Figure \ref{fig:reverse_number}, we verify the numerical accuracy of the estimation of the number of reverse pairs. Given channel output vector $\underline{y}$ and the ordered list of error patterns generated by the CDF-ORBGRAND algorithm, we use \eqref{eq:obtain_D} to identify $\mathcal{D}^{*}$, and then compare the estimated value of $\mathcal{I}$ obtained by \eqref{eq:esti_I} and the exact value of $\mathcal{I}$ obtained by \eqref{eq:inverse_2}. To plot Figure \ref{fig:reverse_number}, we generate $10^6$ independent $\underline{y}$ samples and take the average of the resulting $\mathcal{I}$, with different choices of Eb/N0. The estimated $\mathcal{I}$ is very accurate, with a relative error less than 1\%.

\subsection{Decoding Performance Simulation}\label{subsec:4.2}

\begin{figure}[htbp]
    \centering
    \includegraphics[width=0.45\textwidth]{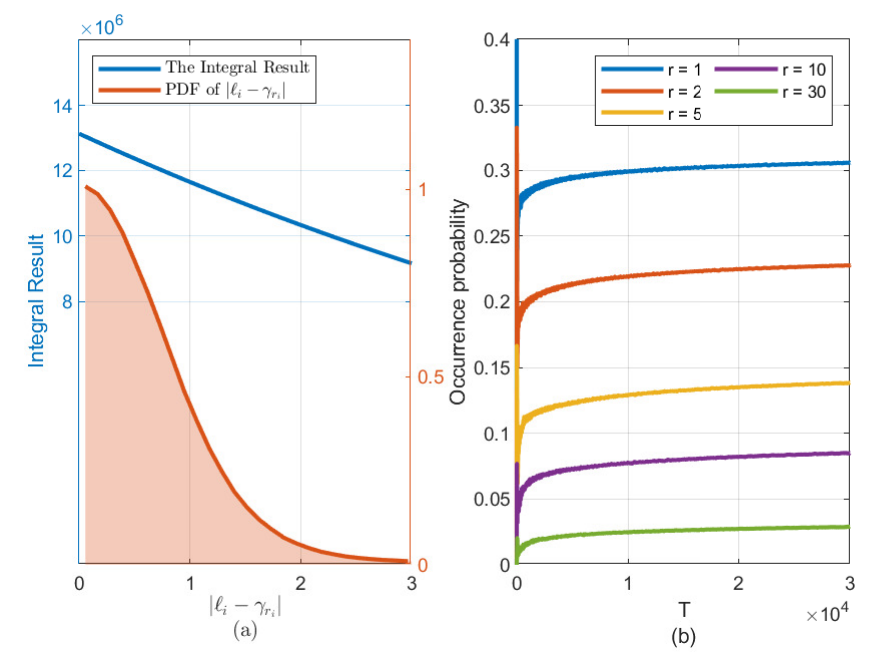}
    \caption{Analyze \eqref{eq:esti_I} under $D=1$. (a): The integral value is insensitive to $|\delta(u,v)|$, the ordinates on the left and right are the integral results and PDF value, the $i$ is considered to be uniformly distributed; (b): The $P_u(T)$ is sensitive to the order of the bits.}
    \label{fig:OandP}
\end{figure}

Figure \ref{fig:OandP} analyzes terms in \eqref{eq:esti_I} when $D = 1$, we set Eb/N0 = 7dB for CDF-ORBGRAND algorithm. The blue line in Figure \ref{fig:OandP} (a) shows the effect of $|\delta(u,v)|$ values (i.e. $\left|\ell_{d_1} - \gamma_{r_{d_1}}\right|$ in \eqref{eq:d_eq_1}) on the integral, its value depends on the received $\underline{\ell}$ and the choice of $d_1$. And the red zone is the PDF of $\left|\ell_{i} - \gamma_{r_{i}}\right|$. It can be seen that in most cases, the integral result is not sensitive to the $|\delta(u,v)|$. Figure \ref{fig:OandP} (b) shows the probability that $e_{i}(t)=1$ in the first $T$ error patterns in $\Tilde{\mathcal{E}}$. The figure shows that bits with smaller ranking values are flipped more frequently, and hence $p_0(T) (1-p_0(T))$ in \eqref{eq:d_eq_1} is very sensitive to the value's order, and bits with smaller $r$ are more likely to be selected.

\begin{figure}[htbp]
    \centering
    \includegraphics[width = 0.45\textwidth]{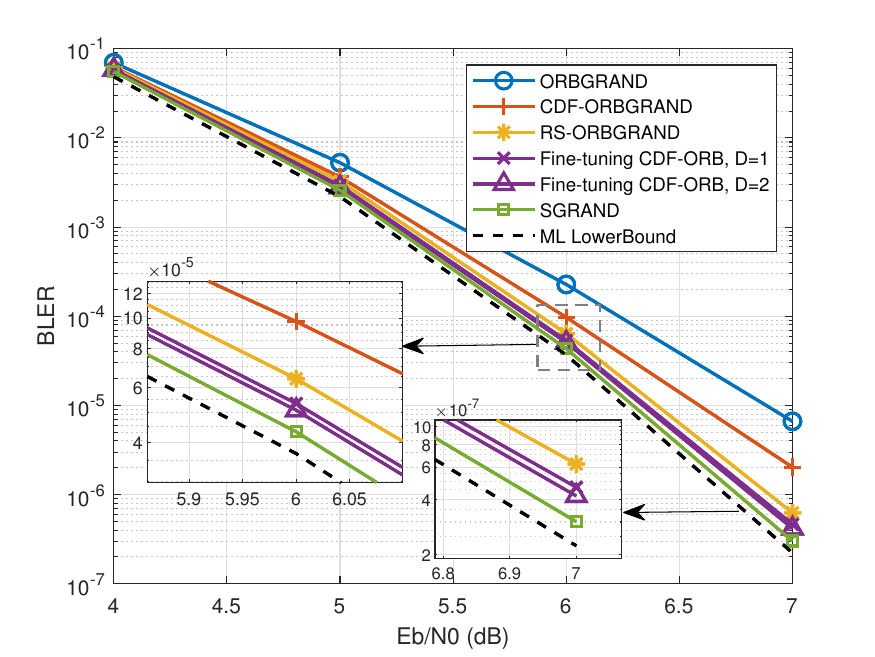}
    \caption{BLER for BCH(127, 113), maximum number of tests $T=10^4$.}
    \label{fig:ftorb}
\end{figure}

In Figure \ref{fig:ftorb}, we compare the BLER for the BCH(127, 113) code decoded by several GRAND algorithms. We conduct fine-tuning for CDF-ORBGRAND, and set $D = 1,2$. For comparison, we include ORBGRAND \cite{duffy2022ordered}, CDF-ORBGRAND without fine-tuning \cite{liu2022orbgrand} \cite{duffy2022ordered}, RS-ORBGRAND \cite{wan2024approaching} which has been recently proposed and achieves the best BLER performance among ORB-type GRAND algorithms, and SGRAND \cite{solomon2020soft}. We also plot the ML lower bound as the ultimate performance limit. We observe from Figure \ref{fig:ftorb} that with fine-tuning, we outperform all ORB-type GRAND algorithms and come very close to SGRAND, using the extra information of only one single exact channel soft value ($D = 1$). And when $D=2$, the improvement is relatively limited.

In Table \ref{tab:2}, we display results of decoding complexity. Due to the adjustment of error patterns, the average number of tests in the presence of fine-tuning is smaller than that of CDF-ORBGRAND. Therefore, although a small amount of additional complexity is required for conducting fine-tuning, its average decoding delay is not higher than that of CDF-ORBGRAND. We also calculate the $\eta$ value of each algorithm with \eqref{eq:e_eta}, and observe that with fine-tuning, the number of reverse pairs is further reduced, compared with CDF-ORBGRAND.

\begin{table}[htbp]
    \centering
    \renewcommand{\arraystretch}{1.0}
    \caption{Number of tests, decoding delay, and $\eta$ metric for different algorithms. The fine-tuning CDF-ORBGRAND with $D = 1$.}
    \label{tab:2}
    \begin{tabular}{|c|c|c|c|c|c|}
        \hline
        & Eb/N0          & 4dB    & 5dB   & 6dB   & 7dB   \\ \hline
        \multirow{5}{*}{\begin{tabular}[c]{@{}c@{}}Average\\ Number\\of Tests\end{tabular}} & ORBGRAND       & 781.6  & 84.05 & 7.076 & 1.469 \\ \cline{2-6} 
        & CDF-ORB   & 721.9  & 65.88 & 5.343 & 1.472 \\ \cline{2-6} 
        & RS-ORB   & 708.2  & 61.91 & 4.512 & 1.382 \\ \cline{2-6} 
        & Fine-tuning CDF & \textcolor{red}{695.8}  & \textcolor{red}{59.02} & \textcolor{red}{4.307} & \textcolor{red}{1.356} \\ \cline{2-6} 
        & SGRAND         & 660.3  & 52.77 & 3.987 & 1.320 \\ \hline
        \multirow{5}{*}{\begin{tabular}[c]{@{}c@{}}Decoding\\ Delay \\(ms)\end{tabular}} & ORBGRAND       & 2.67  & 0.312 & 0.0480 & 0.0273 \\ \cline{2-6} 
        & CDF-ORB   & 2.55  & 0.253 & 0.0406 & 0.0273 \\ \cline{2-6} 
        & RS-ORB   & 2.53  & 0.244 & 0.0381 & 0.0270 \\ \cline{2-6} 
        & Fine-tuning CDF & \textcolor{red}{2.58}  & \textcolor{red}{0.247} & \textcolor{red}{0.0389} & \textcolor{red}{0.0271} \\ \cline{2-6} 
        & SGRAND         & 29.4 & 2.25 & 0.315 & 0.214 \\ \hline
        \multirow{3}{*}{\begin{tabular}[c]{@{}c@{}}Average \\ $\eta$ Value \end{tabular}} & ORBGRAND       & 0.213  & 0.251 & 0.286 & 0.313 \\ \cline{2-6} 
        & CDF-ORB   & 0.188  & 0.211 & 0.232 & 0.233 \\ \cline{2-6} 
        & Fine-tuning CDF & \textcolor{red}{0.171}  & \textcolor{red}{0.191} & \textcolor{red}{0.209} & \textcolor{red}{0.211} \\ \hline
    \end{tabular}
\end{table}

\section{Conclusion}

This paper develops a method for fine-tuning ORB-type GRAND algorithms, by judiciously selecting a small number of exact channel soft values and rearranging the ordered list of error patterns accordingly. The method is based on a well-orderedness metric quantifying the number of reverse pairs in error patterns, and its design is facilitated by leveraging asymptotic results of number partitioning, which provide highly accurate estimates of the number of reverse pairs and the adjustment in fine-tuning. Numerical experiments demonstrate the effectiveness of the proposed fine-tuning method, outperforming ORB-type GRAND algorithms and approaching the ML bound at a negligible overhead in complexity. Admittedly, the derivation of the fine-tuning method has been largely heuristic in nature, and in future work a more solid and rigorous theoretic investigation is expected.

\bibliographystyle{IEEEtran}
\bibliography{Ref}

\end{document}